\documentclass[aps,prb,twocolumn,letterpaper,superscriptaddress,showpacs]{revtex4}

\usepackage{graphicx}
\usepackage{dcolumn}
\usepackage{bm}

\begin{document}

\title{Carrier-mediated ferromagnetic ordering in Mn ion-implanted p$^{+}$GaAs:C}

\author{Y.D. Park}

\email{parkyd@phya.snu.ac.kr}

\affiliation{CSCMR and School of Physics, Seoul National
University, Seoul 151-747 Korea}

\author{J.D. Lim}

\affiliation{School of Physics, Seoul National University, Seoul
151-747 Korea}

\author{K.S. Suh}
\author{S.B. Shim}

\affiliation{CSCMR and School of Physics, Seoul National
University, Seoul 151-747 Korea}

\author{J.S. Lee}

\affiliation{School of Physics, Seoul National University, Seoul
151-747 Korea}

\author{\\C.R. Abernathy}
\author{S.J. Pearton}

\affiliation{Department of Materials Science and Engineering,
University of Florida, Gainesville, Florida 32605}

\author{Y.S. Kim}
\author{Z.G. Khim}

\affiliation{School of Physics, Seoul National University, Seoul
151-747 Korea}

\author{R. G. Wilson}

\affiliation{Consultant, Stevenson Ranch, California 91381}


\date{\today}

\begin{abstract}
Highly p-type GaAs:C was ion-implanted with Mn at differing doses
to produce Mn concentrations in the 1 - 5 at.{\%} range. In
comparison to LT-GaAs and n$^{+}$GaAs:Si samples implanted under
the same conditions, transport and magnetic properties show marked
differences. Transport measurements show anomalies, consistent
with observed magnetic properties and with epi- LT-(Ga,Mn)As, as
well as the extraordinary Hall Effect up to the observed magnetic
ordering temperature (T$_{\text{C}}$). Mn ion-implanted
p$^{+}$GaAs:C with as-grown carrier concentrations $>$10$^{20}$
cm$^{-3}$ show remanent magnetization up to 280 K.
\end{abstract}

\maketitle

Observation of ferromagnetic ordering in highly Mn doped
InAs\cite{Munekata:1989} and GaAs\cite{DeBoeck:1996} has spurred
renewed interest in diluted magnetic semiconductor (DMS) systems
for the possible realization of spintronic devices, ideally
requiring a material system with spin-polarized carriers
compatible with existing semiconductor
electronics.\cite{Wolf:2001} Since the original reports of
magnetic ordering temperature (T$_{\text{C}})$ of 110K for low
temperature molecular beam epitaxy (LT-MBE) prepared (Ga,Mn)As,
researchers elsewhere have reported increases in T$_{\text{C}}$
through optimization of growth conditions and annealing
processes.\cite{Hayashi:2001,Ku:2003,Edmonds:2002} Yet, for
realization of practical devices, a material with T$_{\text{C}}$
near or above room temperature would be desirable. From
theoretical treatment and experimental evidence, carrier
concentration (\emph{p}) plays an important role in mediating
ferromagnetic ordering between localized spins of Mn impurities in
the GaAs
matrix.\cite{Berciu:2001,Kaminski:2002,Korzhavyi:2002,Jungwirth:2002}
Although measurement of \emph{p} from Hall Effect measurements is
complicated by the intrinsic extraordinary Hall Effect
(EHE),\cite{Ohno:1999} \emph{p} measured is far below that of
expected if all Mn acceptors are electrically active (\emph{p} as
low as 15 - 30{\%} of incorporated Mn). Accordingly, only a
fraction (as low as 1/7 of Mn reported by Ohldag \textit{et al}.
from MCD studies\cite{Ohldag:2000}) of the Mn impurities are
experimentally observed to participate magnetically. It is widely
thought that due to the low temperature ($<$300\r{ }C) of the
substrate during growth, total free \emph{p} is compensated by
deep-level donor defects such as As anti-site (As$_{\text{Ga}})$.

The importance of free hole carrier concentration (\emph{p}) has
been demonstrated experimentally by co-doping Sn, a donor
impurity, and Mn in GaAs during growth.\cite{Satoh:2001} As free
hole carriers are compensated, Satoh \textit{et al.} observed a
directly related decrease in T$_{\text{C}}$. Other than co-doping,
modulations of carriers by electric field directly correlate to an
increase and decrease in T$_{\text{C}}$ in (In,Mn)As
(\textit{Ref.~}\onlinecite{Ohno:2000}) and MnGe
(\textit{Ref.~}\onlinecite{Park:2002}). Recent experiments in
post-growth thermal treatments showed markedly higher
T$_{\text{C}}$'s, which increase is thought to be related to an
increase in \textit{p} by decreasing the number of deep level
donor defects, as well as increase in
Mn$_{\text{Ga}}$.\cite{Hayashi:2001,Ku:2003,Edmonds:2002} In
theoretical treatments of ferromagnetic ordering in III-Mn-As, it
is thought that an increase in \textit{p} may directly correspond
to an increase in T$_{\text{C}}$ up to and beyond room
temperature.\cite{Jungwirth:2002} Here, we present structural,
magnetic, and transport properties of high carbon doped GaAs
(p$^{+}$GaAs:C) ion-implanted with Mn. High carbon doping
concentration has been well studied for the development of the
base region in high speed heterojunction bipolar transistors
(HBT).\cite{Abernathy:1992} Under some growth conditions, it is
energetically favorable for carbon to occupy the Arsenic site
(whereas Mn is known to occupy the Ga site) where it acts as a
shallow acceptor with nearly all of the carbon
activated.\cite{Li:1998} Due to a low diffusion coefficient of
carbon in GaAs,\cite{Schubert:1992} carbon concentrations greater
than 10$^{21}$ cm$^{ - 3}$ have been reported, nearly an order of
magnitude higher than \emph{p} reported in LT-(Ga,Mn)As. Carbon
ionization energy in GaAs (E$_{\text{A}}$ - E$_{\text{v}}$) is
nearly half of that of Mn, possibly allowing to independently
control carrier and magnetic impurity concentrations and to
investigate conditions where \emph{p} exceeds the magnetic
impurity concentration, not possible in LT-(Ga,Mn)As.

Epitaxial p$^{+}$GaAs:C (\emph{p} $\sim 3 \times 10^{20}$ cm$^{ -
3} (\sim$ 1.4 {\%} C$_{\text{As}}$), 500 nm thick) films were
grown on semi-insulating (SI) GaAs by gas source molecular beam
epitaxy (GSMBE).\cite{Abernathy:1995} Under these conditions,
co-doping of Mn without significant formation of intermetallic
clusters would be difficult. Introduction of dopants physically by
means of ion-implantation has been well studied to survey various
semiconductors and oxides for DMS,\cite{Pearton:2003} and recently
Scarpulla \textit{et al}. report of single phase (Ga,Mn)As from Mn
ion-implantation and subsequent pulsed-laser melting of
\textit{SI} GaAs with T$_{\text{C}} \sim  80$
K.\cite{Scarpulla:2003} Ion implantation is capable of introducing
dopant concentrations above the usual equilibrium solid solubility
limit. Thus after growth, samples were ion-implanted with Mn at
250 keV with the sample held at $\sim$ 350\r{ }C to minimize
amorphization and with Mn doses of 1, 2, 3, 5 $\times$ 10$^{16}$
cm$^{ - 2 }$(sample A, B, C, D respectively), which corresponds to
nearly 1 - 5 atm.{\%} of Mn. For comparison, LT-MBE prepared GaAs
(LT-GaAs) and high Si doped GaAs (n$^{+}$GaAs:Si) epi-films were
implanted under the same conditions with Mn dose of 3 $\times$
10$^{16}$ cm$^{ - 2}$ (sample X, Y). Details of the implantation
process are given elsewhere.\cite{Pearton:2003} To further
minimize formation of known secondary phases of Ga, Mn, and As, no
post implant anneal was performed.

After ion-implantation, depth profile Auger Electron Spectroscopy
(AES) measurements show Mn to be present down to 300 nm below the
surface and consistent with doping profiles found in previous
studies.\cite{Pearton:2003} To study possible segregation of
implanted species at the surface, electron microprobe x-ray
analysis (EMPA-JEOL JXA-8900R) indicate a homogeneous surface
within resolution of the instrument. High resolution x-ray
diffraction (HRXRD) measurements of as-grown and as-implanted
samples show similar results, with identifiable peaks that can be
only associated to the epi-film and substrate.  The resulting
implanted samples HRXRD measurements do not show possible
secondary phases (such as MnAs and GaMn) or trends as studied by
Moreno \textit{et al}. (by annealing
LT-(Ga,Mn)As).\cite{Moreno:2002} In addition to HRXRD, high
resolution cross-sectional transmission electron microscopy
(HRXTEM) were performed on samples A and D, but due to the
expected high concentrations of structural dislocations, neither
qualitative nor quantitative analysis of secondary precipitates
was possible.

\begin{figure}

\includegraphics{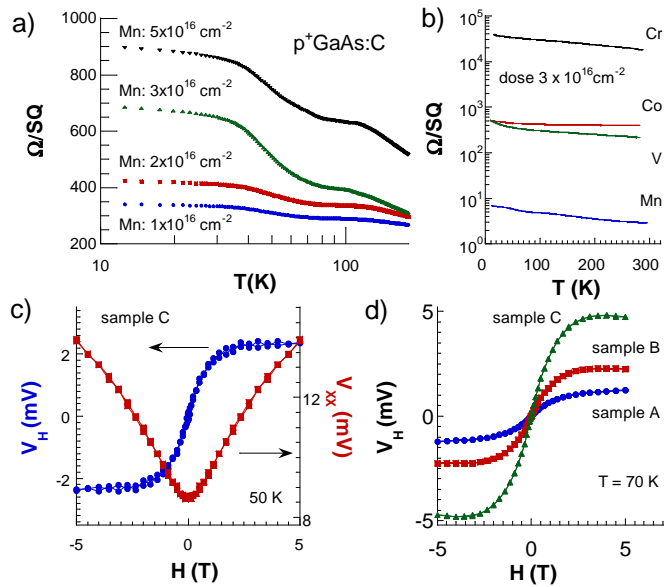}
\caption{\label{fig1}a) Sheet resistance($\Omega/\text{SQ}$) as
function of temperature (T) for samples A - D. b)
$\Omega/\text{SQ}$ as function of T for Co, Cr, Mn, V implanted
p$^{+}$GaAs:C with 3 $\times$ 10$^{16}$ cm$^{ - 2}$ dose. c) AC
magneto-transport measurement of sample C at 50 K with excitation
current of 100 $\mu$A. d) AC Hall Effect measurement of samples A
- C at 70 K. }

\end{figure}

In previous studies of DMS, physical characterization methods such
as HRXRD and HRXTEM as well as others by themselves cannot
completely rule out the presence of secondary phases. Transport
properties, particularly EHE and anomalies near T$_{\text{C}}$, as
well as magnetic properties may be more sensitive and informative
concerning existence of secondary ferromagnetic phases. Hayashi
\textit{et al}. in reporting increase in T$_{\text{C}}$ after
thermal treatment of as-grown (Ga,Mn)As found that even as-grown
samples (with T$_{\text{C}}\sim $ 40K) show a characteristic
`hump' in the resistivity as a function of temperature
plots.\cite{Hayashi:2001} Akinaga \textit{et al}. in studying
nano-magnetic MnAs clusters embedded in GaAs report of
characteristic changes in the slope of resistivity as a function
of temperature curve around 50K, independent of whether clusters
are formed, and assigned the anomaly to a Ga-Mn-As complex in the
matrix.\cite{Akinaga:1998} Standard four point probe DC transport
measurements from 10 K to 300 K were performed using In soldered
contacts in a closed-cycle dewar (Fig.~\ref{fig1}a). Remarkably,
the Mn implanted p$^{+}$GaAs:C samples show a similar features to
LT-MBE prepared (Ga,Mn)As samples. Although a full
metal-to-insulator transition was not observed in all implanted
samples, it is surprising that such features are distinct given
the expectation that the transport properties would be dominated
by damage incurred during the implantation process.

For comparison, as-grown p$^{+}$GaAs:C shows metallic-like
behavior with no distinct features; while samples X and Y show
expected insulator-like behavior due to implantation damage.
Similar insulator-like behavior was observed for Co, Cr, and V
implanted samples into p$^{+}$GaAs:C (Fig.~\ref{fig1}b).
AC-transport measurements using Quantum Design Physical Property
Measurement System (PPMS) (excitation current of 100 $\mu$A at
17.1 Hz) for the temperature range considered (5K to 300K) and
applied magnetic fields up to 5 T indicate positive
magneto-resistance (MR). In granular hybrid systems where
nano-sized transition metal based ferromagnetic intermetallics are
embedded in a semiconductor matrix such as (MnAs:GaAs
(\textit{Ref.}~\onlinecite{Akinaga:1998}); ErAs:GaAs
(\textit{Ref.}~\onlinecite{Schmidt:1999}); and
Mn$_{11}$Ge$_{8}$:Ge (\textit{Ref.}~\onlinecite{Park:2001})), a
cross-over in sign of MR from positive at higher temperatures to
negative at lower temperatures was observed and attributed to
variable hopping mechanisms. Although LT-(Ga,Mn)As show similar
behavior below T$_{\text{C}}$ with a pronounced background
negative MR, a positive near parabolic MR behavior around H = 0
appears for metallic samples below
$\sim$T$_{\text{C}}$.\cite{Iye:1999}

\begin{figure}

\includegraphics{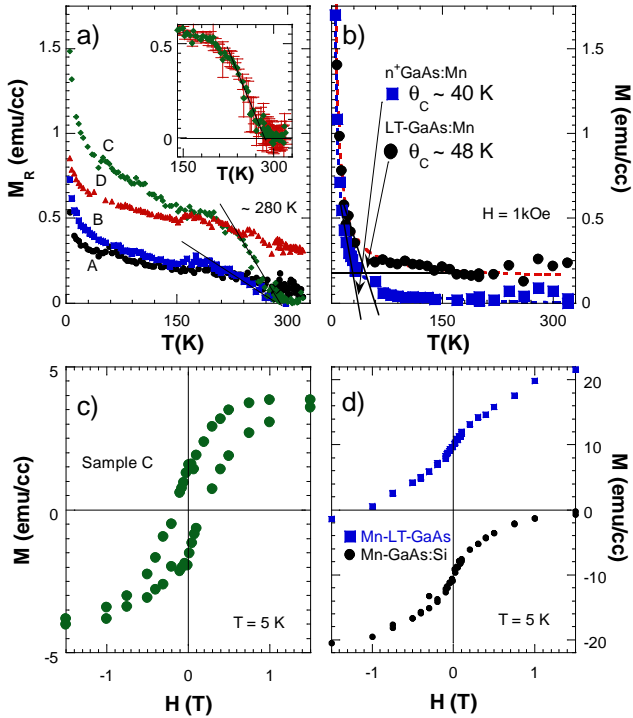}
\caption{\label{fig2}a) M$_{\text{R}}$ as a function of
temperature (T) for Mn implanted p$^{+}$GaAs:C (near
T$_{\text{C}}$ for Sample C, inset). b) Magnetization
($\textbf{M}$) as a function of T for sample X and Y with
Curie-Weiss Law fit. $\textbf{M}$ as a function of applied field
at 5 K for sample C (c) and sample X and Y, offset for clarity
(d).}

\end{figure}

\begin{figure}

\includegraphics{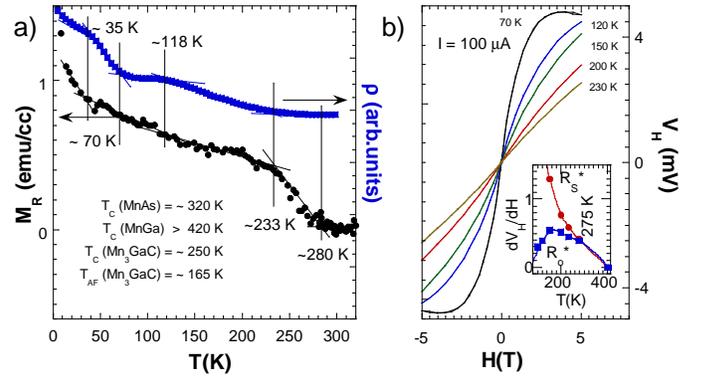}
\caption{\label{fig3}a) M$_{\text{R}}$ and AC resistivity (I = 100
$\mu $A) as a function of temperature for sample C (T$_{\text{C}}
\sim $ 280 K) at H = 0. Anomalies in transport properties
correspond to magnetic properties suggesting changes in
resistivity are due to magnetic ordering in the sample. b) AC-Hall
measurement at various temperature for sample C. Inset shows low
field and high field fit to the Hall response corresponding to
ordinary and extraordinary Hall coefficient for Sample C,
indicating non-linear Hall response below $\sim$275 K.}

\end{figure}

Magneto-transport measurements were carried out to estimate sheet
carrier concentration (\emph{p$_{\text{s}}$}) and determine sheet
resistance($\Omega/\text{SQ}$) using the Van der Pauw geometry.
General trend shows implanted species to enter the GaAs matrix
with an accompanying increase in \emph{p$_{\text{s}}$}
(\emph{p$_{\text{s}}$} = 1.6 $\times$ 10$^{16}$ cm$^{ - 2}$ for
sample A to \emph{p$_{\text{s}}$} = 1.1 $\times$ 10$^{17}$ cm$^{ -
2}$ for sample D at 300 K), and direct correlation to dose and
$\Omega/\text{SQ}$ due to implantation damage. For example, Cr is
a known deep-level donor in GaAs. From Hall measurements, we found
the p$^{+}$GaAs:C Cr ion-implanted sample to be fully compensated
and n-type (\emph{n$_{\text{s}}$} = 1.1 $\times$ 10$^{14}$ cm$^{ -
2}$ at 300 K). In previous reports of ferromagnetic ordering in
single phase LT-(Ga,Mn)As, the observation of EHE has been
attributed to spin-polarized carriers that mediate ferromagnetic
ordering between localized spins.\cite{Jungwirth:2002_2} AC Hall
measurements for samples A - C are plotted in Fig.~\ref{fig1}d,
and the behavior is consistent with previously reported
LT-(Ga,Mn)As with onset of non-linear Hall response below
$\sim$280 K. Further details of the magneto-transport measurements
will be presented elsewhere.

From the transport measurements, it is highly probable that
significant levels of the implanted species are electrically
active at room temperature and that for all samples, implantation
damage dominates transport properties (show insulator-like
behavior) except that of Mn implanted p$^{+}$GaAs:C, which shows
changes in slope as reported in LT-(Ga,Mn)As samples, in which
anomalies coincide near the magnetic transition temperatures. This
feature is absent in ion-implanted LT-GaAs and n$^{+}$GaAs:Si
samples. Although transport measurements were not reported,
Theodoropoulou \textit{et al}. report of unconventional
carrier-mediated ferromagnetism in ion-implanted
(Ga,Mn)P:C.\cite{Theodoropoulou:2002} They showed a distinct
difference in magnetic properties between Mn ion-implanted GaP:C
and n-GaP, consistent with hole mediated ferromagnetic ordering in
III-V DMS.

The magnetic properties of Mn implanted GaAs:C, LT-GaAs, and
GaAs:Si was measured using Quantum Design Magnetic Property
Measurement System (MPMS). Figure \ref{fig2}a plots the remanent
magnetization (M$_{\text{R}}$) as a function of temperature for
samples A - D. Magnetic field of 5 Tesla was applied at 5 K and
switched off, followed by series of magnetization measurements at
zero applied field and at various temperatures (up to 320K). For
samples A - C, non-zero M$_{\text{R}}$ was found for temperatures
below $\sim$ 280 K. Unlike LT-(Ga,Mn)As, T$_{\text{C}}$'s of
samples A - C are weakly dependent on Mn content, to be discussed
later. Similar measurement of samples X and Y show a near zero
flat response indicating paramagnetic-like behavior, which is
confirmed by magnetic hysteresis measurements at 5 K with near
zero M$_{\text{R}}$ (Fig.~\ref{fig2}d). Magnetization as a
function of temperature (M vs.~T) measurements with applied field
of 1000 Oe show Curie-Weiss temperatures ($\theta _{\text{C}}$)
below $\sim $50 K for samples X and Y (Fig. 2b), which is
confirmed by equal traces for zero field cooled and field cooled
measurement with a read field of 100 Oe from 5 K - 300 K. The B-H
loops for p$^{+}$GaAs:C implanted samples show well-defined
magnetic hysteresis loops with high M$_{\text{R}}$ compared to
GaAs:Si and LT-GaAs implanted samples (Fig.~\ref{fig2}c\&d).

As reported for LT-(Ga,Mn)As samples on the insulator side of
metal-insulator transition, complete magnetic saturation in
samples A - C was found to be difficult; thus, calculation of
magnetic moment per Mn atom would yield incomplete values.  For
sample C, at 5 K and at M$_{\text{R}}$, we approximate less than
$\sim$ 1/10 of the implanted Mn to magnetically contribute,
assuming Mn spin S = 5/2 and the Land\'{e} factor
g$_{\text{Mn}}=2$. From B-H loops of p$^{+}$GaAs:C implanted
samples, we found the coercive field (H$_{\text{C}}$) to range
from few hundred Oersteds to ~2000 Oe with maximum coercive field
measured from sample C. A similar trend for approximate magnetic
saturation (M$_{\text{S}}$) was found with maximum $\sim$
M$_{\text{s}}$ corresponding to sample C with Mn dose of 3$\times
10^{16}$ cm$^{-2}$.

The weak dependence of T$_{\text{C}}$ to Mn dose may suggest an
inhomogeneous profile (i.e., equal peak concentration at some
distance from surface for samples A - C). As depth profile AES
showed a near constant Mn concentration, we note that \emph{p} is
greater than the effective magnetic impurity concentration
($x_{\text{eff}}$N$_{0}$) for samples A - C unlike LT-(Ga,Mn)As
where \emph{p} is less than Mn concentration. In their study of
carrier-induced ferromagnetism in p-ZnMnTe, Ferrand \textit{et
al.} propose for the case where \emph{p} $>$
$x_{\text{eff}}$N$_{0}$ that the Rudermann-Kittel-Kasuya-Yosida
(RKKY) model best describes the ferromagnetic orderingm since the
Zener model ceases to be valid.\cite{Ferrand:2001} In such a view,
the weak dependence of T$_{\text{C}}$ on implanted dose may be
explained as the tendency of T$_{\text{C}}$ to increase with
$x_{\text{eff}}$N$_{0}$ being offset Mn-Mn interactions as
\emph{p} increases in this RKKY regime($p>x_{\text{eff}}$N$_{0}$).

Again, without the observed unexpected differences in magnetic
properties of Mn implanted p$^{+}$GaAs:C samples with
n$^{+}$GaAs:Si and LT-GaAs implanted samples, observed magnetic
properties, including possibly the observed weak dependence of
T$_{\text{C}}$ to Mn implantation does, might be easily assigned
to intermetallic ferromagnetic precipitates such as MnAs, GaMn,
and Ga-Mn-As. For the samples considered, sample X would be most
susceptible to formation of ferromagnetic precipitates with
well-known excess of As in LT-GaAs. In their careful study of
different possible secondary phases, Moreno \textit{et al.} have
identified three possible precipitates: hexagonal MnAs, Zinc
Blende Mn(Ga)As, and MnGa.\cite{Moreno:2002} For samples A - C,
ion implanted p$^{+}$GaAs:C samples, the T$_{\text{C}}$ and
magnetization values measured do not correspond to the mentioned
precipitates as well as Mn$_{3}$GaC (T$_{\text{C}}$ of $\sim$250
K), to be detailed later.\cite{Bouchaud:1966} From M$_{\text{R}}$
as a function of temperature trace, sample D behavior of apparent
T$_{\text{C}}$ greater than 320 K points to formation of MnGa
(T$_{\text{C}}$ $>$ 400 K) precipitates in agreement with Moreno
\textit{et al.} and Shi \textit{et al}.\cite{Shi:1996} The
magnetic properties of sample A - C corroborate our findings from
HRXRD measurements.

In comparing the magnetic properties of samples A - C with Mn
ion-implanted sample X and Y, it is evident that high hole carrier
concentrations mediate ferromagnetic ordering between localized
spins. Whether these localized spins are associated with
substitutional Mn$^{2 + }$ ions or a physically undetected
intermetallic ferromagnetic clusters, there is a strong evidence
that free carrier concentration play an important role in
mediating this ferromagnetic ordering, evident in increased
M$_{\text{R}}$ and T$_{\text{C}}$, especially comparing carbon
doped GaAs implanted samples (A - C) to LT-GaAs and Si doped GaAs
samples (X \& Y). From physical characterization studies, we did
not observe any secondary phases or trends observed by others
studying intermetallic clusters in semiconductor matrix for all
samples except with highest Mn dose (sample D). In addition, if
secondary ferromagnetic phases created by the implanted ions were
responsible for the observed magnetic properties, then we expect
similar results in Mn ion-implanted carbon doped p$^{+}$GaAs as
well as in LT-GaAs and Si doped n$^{+}$GaAs, as formation of such
physically undetected ferromagnetic phases as MnAs, GaMn, and
Mn(Ga)As would be, at the least, equally probable in all samples
considered. For possible undetected ferromagnetic phases unique to
carbon doped samples, Mn$_{3}$GaC with ferromagnetic transition
temperature $\sim$250 K is a possibility, but this perovskite-type
material has a well known anti-ferromagnetic transition at
$\sim$165 K at zero magnetic field, which should have been clearly
evident from the M$_{\text{R}}$ vs. T measurements as well as M
vs. H measurements at 5 K.\cite{Bouchaud:1966} From transport
measurements, we observed characteristic anomalies corresponding
to magnetic properties as well as the EHE (Fig.~\ref{fig3}), a
telling-sign that the carriers are spin-polarized and mediate
ferromagnetic ordering between localized spins. Plotting Hall
Effect response (dV$_{\text{H}}$/dH) at high fields (directly
proportional to the ordinary Hall Effect coefficient
(R$_{\text{o}}$)) and at low fields (EHE coefficient
(R$_{\text{S}}$)), we observe a distinct difference near
T$_{\text{C}}$ observed from temperature dependence of
M$_{\text{R}}$ (Fig.~\ref{fig3}b inset).

To summarize, we have observed remanent magnetization up to
$\sim$280 K for Mn ion-implanted p$^{+}$GaAs:C.  From physical
property measurements, we cannot attribute the magnetic properties
to an observed presence of secondary ferromagnetic phases such as
MnAs, Mn(Ga)As, Mn$_{3}$GaC, and GaMn.  This result is supported
by magnetic properties of LT-GaAs and n$^{+}$GaAs:Si Mn
ion-implanted (under same conditions) samples, which show neither
the marked increase in magnetic ordering temperatures and remanent
magnetization nor magnetic properties that can be attributed to
physically undetected ferromagnetic secondary phases.  This
difference between samples (Mn implanted carbon doped p$^{+}$GaAs
vs. Mn implanted LT-GaAs and Si doped n$^{+}$GaAs) points to a
role of the carbon acceptor impurity, which substitutionally
prefers the group V (or As) site and provide hole carriers.
Whether this role of carbon acceptor in the observation of
increased magnetic properties is due to increased number of free
carrier concentration or due to a secondary ferromagnetic phase
unique to carbon (Mn$_{3}$GaC) arguably cannot be currently
irrefutably answered in detail.  But, such properties as
antiferromagnetic transition attributable to Mn$_{3}$GaC were not
observed. Transport measurements of Mn ion-implanted p$^{+}$GaAs:C
samples do not agree with results as reported by others of
nanometer-sized ferromagnetic clusters embedded in semiconductor
matrix.  Rather transport measurements correspond with temperature
dependence of magnetic properties, much like LT-(Ga,Mn)As.
Additionally, observation of non-linear Hall response (or EHE)
corresponds to observed magnetic ordering temperature.

We would like to thank Dr. J. Chang of KIST with the TEM analysis;
and Drs. H.C. Kim  and S.-H. Park of KBSI Material Science
Laboratory with access to PPMS. The work at SNU was supported by
KOSEF and Samsung Electronics Endowment through CSCMR.  The work
at UF was supported by NSF DMR0101438 and ECS 02242303 and by ARO
under DAAD 190110701 and 19021420.

\end{document}